\documentclass[aps,pre,showpacs,preprintnumbers,twocolumn,nofootinbib,superscriptaddress]{revtex4-1}

\usepackage{geometry}                
\usepackage{graphicx}
\usepackage{amssymb}
\usepackage{dsfont}
\usepackage{epstopdf}
\usepackage{color}
\usepackage{mathtools}

\definecolor{red}{rgb}{1,0,0}
\definecolor{blue}{rgb}{0,0,1}
\definecolor{black}{rgb}{0,0,0}
\newcommand{\blue}{ }

\newcommand{\p}{\partial}
\newcommand{\eq}[1]{\begin{align}#1\end{align}}
\newcommand{\eqs}[1]{\begin{align*}#1\end{align*}}
\newcommand{\ffrac}[2]{\mbox{$\frac{#1}{#2}$}}
\newcommand{\half}{\mbox{$\frac{1}{2}$}}
\newcommand{\OO}{\mathcal{O}}
\newcommand{\tr}{\text{tr}}

\newcommand{\al}{|\lambda|}

\usepackage{calrsfs}

\usepackage{scalerel,stackengine}
\stackMath
\newcommand\widecheck[1]{%
\savestack{\tmpbox}{\stretchto{%
  \scaleto{%
    \scalerel*[\widthof{\ensuremath{#1}}]{\kern-.6pt\bigwedge\kern-.6pt}%
    {\rule[-\textheight/2]{1ex}{\textheight}}
  }{\textheight}%
}{0.5ex}}%
\stackon[1pt]{#1}{\scalebox{-1}{\tmpbox}}%
}

\newcommand{\PP}{\mathbb{P}}

\begin{document}
\title{Breakdown of random matrix universality in Markov models}
\author{Faheem Mosam}
\affiliation{Department of Physics, Ryerson University, M5B 2K3, Toronto, Canada}
\author{Diego Vidaurre}
\affiliation{Department of Psychiatry, Oxford University, UK}
\affiliation{Center for Functionally Integrative Neuroscience, Department of Clinical Medicine, Aarhus University, Denmark}
\author{Eric De Giuli}
\affiliation{Department of Physics, Ryerson University, M5B 2K3, Toronto, Canada}

\begin{abstract}
Biological systems need to react to stimuli over a broad spectrum of timescales. If and how this ability can emerge without external fine-tuning is a puzzle. We consider here this problem in discrete Markovian systems, where we can leverage results from random matrix theory. Indeed, generic {\it large} transition matrices are governed by universal results, which predict the absence of long timescales unless fine-tuned. We consider an ensemble of transition matrices and motivate a temperature-like variable that controls the dynamic range of matrix elements, which we show  plays a crucial role in the applicability of the large matrix limit: as the dynamic range increases, a phase transition occurs whereby the random matrix theory result is avoided, and long relaxation times ensue, in the entire `ordered' phase. We furthermore show that this phase transition is accompanied by a drop in the entropy rate and a peak in complexity, as measured by predictive information (Bialek, Nemenman, Tishby {\it Neural Computation} 13(21) 2001). Extending the Markov model to a Hidden Markov model (HMM), we show that observable sequences inherit properties of the hidden sequences, allowing HMMs to be understood in terms of more accessible Markov models. We then apply our findings to fMRI data from 820 human subjects scanned at wakeful rest. We show that the data can be quantitatively understood in terms of the random model, and that brain activity lies close to the phase transition when engaged in unconstrained, task-free cognition -- supporting the brain criticality hypothesis in this context. 
\end{abstract}
\maketitle

\section{Introduction}
\subsection{Background \& Motivation}

Complex systems typically display a broad spectrum of relaxation timescales \cite{Munoz18}. This phenomenon is often linked to proposed critical behavior \cite{Mora11,Munoz18}, in analogy with equilibrium statistical mechanics. However, equilibrium systems must be tuned to a critical point. For example, a ferromagnetic material has a phase transition between paramagnetism and ferromagnetism at the Curie temperature $T_C$. The system temperature $T$ must be externally tuned to the vicinity of $T_C$ in order to see the long-range scale-free correlations and broad relaxations characteristic of a critical point. {\blue Since there is ample evidence for critical behavior in out-of-equilibrium complex systems, we must either find a generic mechanism for self-tuning to a critical point, multiple context-dependent mechanisms for criticality, or challenge the equilibrium analogy entirely. 

In \cite{Bak87}, Bak, Tang, and Weisenfeld proposed a simple mechanism for the ubiquity of critical behavior in slowly driven systems. In particular, they showed how in the limit of asymptotically slow driving compared to internal relaxation, generic systems can show scale-free behavior, thus mimicking equilibrium systems at a critical point. }
 While indeed generic, this theory does not predict different universality classes that would distinguish various complex systems, characterized for example by critical exponents \cite{Chaikin00}. It is also of questionable relevance to biological systems, where driving is often not slow compared to response. {\blue Thus, even if the evidence of criticality in complex systems is accepted, the means by which criticality is obtained is still unclear. 
 
Recent work has instead sought context-specific mechanisms for self-organized criticality, particularly in the brain \cite{Levina07,Zeraati20} and in glassy systems \cite{Muller14}. While of clear relevance in their particular domains, these works do not address the question of generic emergence of long time scales. 

Here we study this problem in the simple but general setting of discrete Markov models. Recent work has focussed on the eigenvalue spectra of Markov models, most often in the symmetric case \cite{Kuhn15,Kuhn15a}, on first-passage times \cite{Bartolucci21}, and on general properties of Markov ensembles \cite{Horvat09,Oliveira19}. Here, we define an appropriate ensemble of Markov models that spans the full dynamic regime from systems that just produce random noise, to those that are fully deterministic. Although we make use of previous results from random matrix theory \cite{Girko85,Tao08,Bordenave12}, our ensemble is distinct from those considered previously \cite{Horvat09,Oliveira19}. }


We show that these regimes are separated by a transition with many aspects of a phase transition, whose location can be identified by a breakdown of large-N random matrix theory results. We examine the phases in terms of network- and information-theoretic quantities, and show in particular that complexity peaks near the transition. We then extend these results to Hidden Markov models (HMMs), in which a hidden Markovian system outputs `observable' sequences. Despite being more difficult to analytically characterize, the HMM inherits many properties of the original simple Markov model, which aids its interpretation. Finally, we consider neural fMRI data from \cite{Vidaurre17} that was previously fitted to a HMM. We find that the fMRI data lie very close to the phase transition as predicted analytically. This can be interpreted as support of the brain criticality hypothesis \cite{Arcangelis06,Levina07,Beggs08,Deco12,Hesse14,Tkacik15,Fontenele19,Zeraati20,Fosque21}. \\

\subsection{Continuous-time dynamics: } To show why prediction of a broad spectrum of timescales is challenging, and to motivate our model, we consider first a generic continuous-time linear dynamics that is often applied to neural data \cite{Chen20}:
\eq{ \label{x1}
\p_t x_i = -x_i + \tilde M_{ij} x_j + \xi_i,
}
where $x_i$ is a dynamical variable, $\tilde M_{ij}$ an interaction matrix, and $\xi_i$ noise, with $i=1,\ldots,n$. For example, \eqref{x1} is a crude model for a neural network, where the $x_i$ could be the firing rate of the i$^{th}$ neuron. The first term on the right-hand side is a stabilizing term, that will send $x_i \to 0$ in the absence of any external stimuli. In the normalization of \eqref{x1}, this occurs over a timescale of order unity. The second term on the right-hand side represents the interactions from other neurons: if $\tilde M_{ij}>0$, then neuron $j$ is excitatory for neuron $i$, while in $\tilde M_{ij}<0$, then neuron $j$ is inhibitory for neuron $i$. The final term $\xi_i$ is a noise term, which could represent input from the external world, for example from the part of the brain that is not explicitly modelled. 

The dynamics of \eqref{x1} depends largely on a balance between the stabilizing effect of the first term, and the interactions that could lead to self-sustained activity. Since \eqref{x1} is linear, this balance is encoded in the spectrum of the matrix $\tilde M$, or more precisely the matrix $-\delta_{ij} + \tilde M_{ij}$ that includes the first two terms on the right-hand side of \eqref{x1}. (Here $\delta_{ij}$ is a Kronecker delta, $\delta_{ij}=1$ if $i=j$ and 0 otherwise).

The relaxation times $\tau_i$ are related to the eigenvalues $\tilde \lambda_i$ of $\tilde M_{ij}$ by \cite{Chen20}
\eq{
\tau_i = \frac{1}{1-\mbox{Re}[\tilde\lambda_i]},
}
where the stability limit $\mbox{Re}[\tilde\lambda_c] = 1$ is { due to the unit coefficient of the stabilizing term in \eqref{x1}. }Generically, the matrix $\tilde M_{ij}$ is not symmetric and has no special symmetries. Then, in the large $n$ limit of practical interest, its spectrum will generically tend towards the Girko law in which the eigenvalues are uniformly spread within a circle in the complex plane, say of radius $r$ \cite{Girko85,Tao08,Bordenave12}. Since relaxation times only become large when the real part of the eigenvalues approaches 1, the system is then either stable, if $r<1$, with a finite maximum time-scale, or unstable, if $r>1$, except precisely at the critical point $r=1$. Thus universality of random matrices seems to be at odds with emergence of long timescales, without fine-tuning \cite{Chen20}.  

In reality, $\tilde M$ is usually not fixed but rather evolves slowly. For example, in neural networks it could represent the strength of synaptic connections. Then we can ask if some dynamics will lead $\tilde M$ to generically show long timescales. It was recently shown in a detailed study of the symmetric case \cite{Chen20} that a regulation mechanism based on the overall activity can lead to an accumulation of long timescales. While such an approach can engage with problem-specific dynamical details, it remains a challenge to identify universal mechanisms in which long timescales emerge.


\begin{figure*}[th!] 
\includegraphics[width=\textwidth]{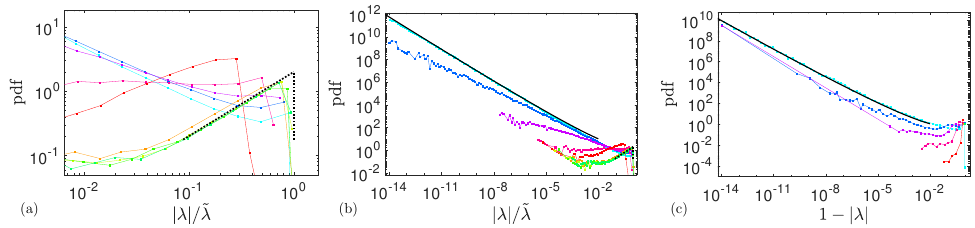}
  \caption[Figure 1]{\label{fig1} Transition matrix spectra for $N=128$ as a function of $\epsilon$, from $\epsilon/\epsilon_c = 10^{2.4}$ (green) through $\epsilon/\epsilon_c = 10^{0}$ (pink) to $\epsilon/\epsilon_c = 10^{-2.4}$ (cyan). $\epsilon$ varies logarithmically by factors of 4; complete labels are shown in Fig 5. (a) Normalized spectrum of eigenvalue magnitude, compared to the random matrix theory prediction $\PP_{RMT}(\al)=2\al, \al <1$. The latter holds for $\epsilon \gtrsim \epsilon_c$. (b) $\PP(\al)$ over a wide range, showing divergent tail $\PP(\al)\sim 1/[|\lambda|\log(1/|\lambda|)^{3/2}]$ (black) at $\epsilon/\epsilon_c = 10^{-2.4}$. (c) Spectrum near $\al=1$. In this regime, at $\epsilon/\epsilon_c = 10^{-2.4}$ we find $\PP(\al)\sim 1/[(1-\al)\log(1/(1-\al))^{5/2}]$ (black).  
  }
\end{figure*}

The fundamental difficulty in finding a generic pathway to long timescales in \eqref{x1} is that the stability limit $\mbox{Re}[\tilde\lambda_c] = 1$ is not universal, but is simply when the interactions determined by $\tilde M$ overwhelm the stabilizing force in \eqref{x1}. In the present work we propose an alternative strategy. Instead of considering the dynamics directly, as in \eqref{x1}, we consider the motion in phase space. That is, instead of considering the evolution of the $\{ x_i(t) \}$, we consider the probability distribution over those variables, $\rho( x_1,x_2,\ldots,x_n; t)$. The latter evolves according to a master equation. Schematically this takes the form
\eq{ \label{markov1}
\rho(y,t+dt) = \sum_x M_{yx} \rho(x,t), 
}
where $M$ is the transition matrix giving the probability for a transition from state $x$ to state $y$, where e.g $x=(x_1,x_2,\ldots,x_n)$ encodes the state of the entire system. \eqref{markov1} is written in a discrete notation, but carries over without difficulty to the fully continuous case. For a derivation of \eqref{markov1} from \eqref{x1} see the Appendix. 

The transition matrix is related to the terms in \eqref{x1}. If each $x_i$ has $n_x$ different states, then $M$ is an $N\times N$ matrix, with $N=(n_x)^n$. Thus for a fully continuous system, $M$ becomes an infinite matrix, an operator. To avoid technical difficulties, we stick here and in the following to the discrete case, where, however, $N$ may be very large. We note that $\sum_{y} M_{yx} = 1$ and each $M_{yx} \geq 0$. Such matrices are called left-stochastic matrices.

The Markovian dynamics \eqref{markov1} is attractive because of its manifest linearity, but it has two serious disadvantages: first, it takes place in an enormous space; and second, the details of the dynamics are hidden inside the definition of $M$. Yet, if our interest is in extracting generic properties of complex systems, we can simply treat $M$ as a random matrix and look for universal properties in the large $N$ limit. This will be the perspective taken here. It is most appropriate in a regime in which the underlying dynamical system has very strong interactions, and is sampled over a timescale much larger than the microscopic timescale. Indeed, under these assumptions, the phase space dynamics will have scrambled local structural correlations arising from continuity of the underlying dynamical process. In some sense, this is the opposite limit to that which favors analysis based on \eqref{x1}.

What is gained from working in phase space? The key distinction between \eqref{x1} and \eqref{markov1} is that while $\tilde M$ was relatively unconstrained, $M$ is instead a stochastic matrix. For simplicity we assume that it is irreducible and aperiodic \footnote{Periodicity complicates the analysis of long-time behavior since there is then no unique stationary state. In our random model, such behaviors are not encountered so there is no need to complicate matters here.}. This entails that $M$ obeys the Perron-Frobenius theorem, in particular its largest eigenvalue has unit modulus. Thus $M$ has a natural scale, unity, induced from probability conservation.  

In this work we consider a generic model for $M$, with a `temperature' scale $\epsilon$ that controls its heterogeneity. We will show that the properties of $M$ change quite dramatically at a certain scale $\epsilon_c$, which in a sense to be explained below, corresponds to the breakdown of large-$N$ random matrix theory. We show that in the entire low-temperature phase, $M$ shows large relaxation times.

In a second part, we promote \eqref{markov1} to a hidden Markov model, where an observable sequence $\{o(t)\}$ is produced by an additional emission matrix $O_{ox}$ that depends on the `hidden' state $x(t)$. We will find that many properties of the observable sequences mirror those of the underlying hidden sequences. This can be useful in practice where one only has direct access to the observable sequences. We will explore in particular how the observable sequences show the signature of the transition. \\

\section{Markov models} 
Markov models arise naturally in a variety of contexts, not necessarily related to any underlying dynamical system such as \eqref{x1}. We consider a system with $N$ states. We can always write $M_{ab}$ in terms of a more primitive matrix $Q_{ab}$ such that $M_{ab}=Q_{ab}/\sum_c Q_{ac}$; then the $Q_{ab}\geq 0$ but do not require any special normalization. 

It has been proved in great generality that $M_{ab}$ has universal features in the large $N$ limit \cite{Girko85,Tao08,Bordenave12}. Let the $Q_{ab}$ be identically and independently distributed, with bounded density, mean $\mu$, and finite variance $\sigma^2$. Then as $N \to \infty$ the spectrum of $M_{ab}$ converges to the uniform law on the disk $|\lambda| < \lambda_c$ in the complex plane (Girko's law), with
\eq{ \label{lamc}
\lambda_c = \sigma/(\mu \sqrt{N}).
}
In practice, this law works well for large but finite $N$, unless a transition intervenes, to be discussed presently.

Solving \eqref{markov1} in terms of the left and right eigenvectors $w_{a,\lambda}$ and $v_{a,\lambda}$ and the stationary right eigenvector $\pi_a = v_{a,1}$, we have
\eq{
\rho_a(t) = \pi_a + \sum_{\lambda \neq 1} A_\lambda \lambda^t v_{a,\lambda},
}
with $A_\lambda = \Sigma_b w_{b,\lambda} \rho_b(0)$. Since $\lambda^t = e^{t \log(\lambda)}$ we see that the relaxation times are
\eq{
\tau_\lambda = -\frac{1}{\log |\lambda|}
}
For sufficiently large $N$, $\lambda_c<1$ and all relaxation times are finite. In this regime, the asymptotic large $N$ random matrix theory (RMT) result applies. What happens when $\lambda_c>1$? Since $M$ is a stochastic matrix, it cannot have eigenvalues with magnitude larger than unity, thus the large-$N$ result must break down, and the spectrum must reorganize. It is this collision of random matrix theory with Perron-Frobenius that is the main subject of this article. 

\begin{figure*}[th!] 
\includegraphics[width=\textwidth]{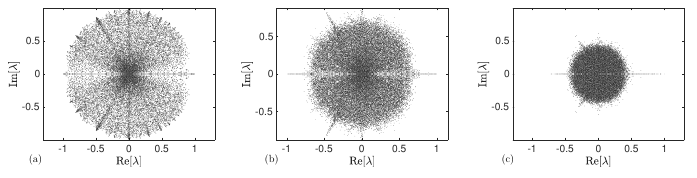}
  \caption[Figure 2]{\label{figspec} Spectra of 1000 random Markov matrices in the complex $\lambda$ plane, with $N=64$, for (a) $\epsilon/\epsilon_c = 0.03$; (b) $\epsilon/\epsilon_c = 0.5$; (c) $\epsilon/\epsilon_c = 2$. 
  }
\end{figure*}

\subsection{Random matrix ensemble} 

Since our aim is to elucidate generic properties of Markov models, we need to identify an appropriate ensemble of the $Q$ matrices. Each element $Q_{ab}$ corresponds to an (unnormalized) transition rate between two states. As a simple model, suppose that the matrix elements $Q_{ab}$ are identically and independently formed from multiplicatively accumulating many factors\footnote{This is the analog of applying the central limit theorem to motivate a Gaussian distribution, but here since the $Q_{ab} \geq 0$, we must multiply factors rather than sum them.}. Then, if the factors are sufficiently independent, $Q_{ab}$ will have a lognormal distribution \cite{Sornette97}, viz.,
\eq{ \label{lognormal}
\PP(Q_{ab}) \propto \frac{1}{Q_{ab}} e^{-\epsilon \log^2(Q_{ab}/q)},
}
{\blue where $\epsilon$ and $q$ are parameters of the distribution. }The normalized fluctuations of $\log Q_{ab}$ have magnitude
\eq{ \label{logfluc}
\frac{\langle Q_{ab}^2 \rangle}{\langle Q_{ab} \rangle^2} - 1 = e^{1/(2\epsilon)} - 1
}
We call $\epsilon$ the temperature of the matrix. {\blue The parameter $q$ controls the overall scaling of the elements via
\eq{
\left\langle \log Q_{ab} \right\rangle = \log q
}
Since $q$ drops out of the $M$ matrix, it will not be varied in what follows. Note that $Q_{aa}$ is sampled from the same distribution as $Q_{ab}$ for $a \neq b$. Since \eqref{lognormal} implies that $Q_{aa} \neq 0$, this also implies that the Markov chain is aperiodic. 

For a matrix $Q_{ab}$ with average $\overline{Q} = \ffrac{1}{N^2}\sum_{a,b} Q_{ab}$, we define its heterogeneity 
\eq{
h(Q) = \frac{1}{N^2} \sum_{a,b} \log^2(Q_{ab}/\overline{Q}),
}
which measures the dispersion of its elements about its mean. }Notice that heterogeneity is invariant if $Q_{ab} \to Q'_{ab} = \overline{Q}^2/Q_{ab}$. It is an appropriate measure for matrices (and tensors \cite{DeGiuli19}) with positive entries.

{\blue Over the ensemble defined by \eqref{lognormal}, we have $\langle \overline{Q} \rangle = q e^{1/(4\epsilon)}$. so that $\langle h \rangle = (2\epsilon)^{-1} + (16 \epsilon^2)^{-1}$. This implies that }`hot' matrices with $\epsilon \gtrsim 1$ have a small heterogeneity, and `cold' matrices with $\epsilon \lesssim 1$ have a large heterogeneity. Intuitively, $\epsilon$ measures how uniformly distributed the transition probabilities are: when $\epsilon \to \infty$, all $Q_{ab} \to \overline{Q}$, while at small $\epsilon$ the $Q_{ab}$ will vary over a huge range. 

Combining \eqref{lamc} with \eqref{logfluc}, we see that $\lambda_c = \sqrt{e^{1/(2\epsilon)}-1}/\sqrt{N}$, thus we predict a critical temperature
\eq{
\epsilon_c(N) = 1/(2\log(N+1))
}
where $\lambda_c = 1$. The circular law cannot apply for $\epsilon<\epsilon_c$. The relevant eigenvalue scale is therefore
\eq{
\tilde\lambda = \mbox{min}(1,\lambda_c),
}
so that for $\epsilon<\epsilon_c, \tilde\lambda=1$. 

This phenomenon is illustrated in Fig. 1a, which shows the distribution of $|\lambda|/\tilde\lambda$ for $N=128$ and a large range of $\epsilon$; qualitatively identical results hold for $N$ down to 16. The RMT result is
\eq{
\PP(|\lambda|) = 2\frac{|\lambda|}{\lambda_c}, \qquad |\lambda|<\lambda_c
}
and holds well for $\lambda > \lambda_{min}$, even at finite $N$, for $\epsilon \gtrsim \epsilon_c$. The lower cutoff $\lambda_{min}$ simply corresponds to the value of $\lambda$ for which there is typically only one eigenvalue left in the disk of radius $\lambda_{min}$, which turns out to be $\lambda_{min}/\lambda_c = 1/\sqrt{N}$. 

More interesting is the inevitable failure of the RMT result for $\epsilon \lesssim \epsilon_c$. We find that when $\epsilon \lesssim \epsilon_c$, $\PP(|\lambda|)$ develops a diverging tail at small $|\lambda|$, shown in Fig. 1b. At the smallest $\epsilon$, our data is consistent with a barely convergent form $\PP(|\lambda|) \sim 1/[|\lambda|\log(1/|\lambda|)^{3/2}]$ over ten orders of magnitude in $|\lambda|$. Any such form implies that as the matrices increase in size, arbitrarily small values of $|\lambda|$ will be found. Assuming this form, the smallest magnitude eigenvalue in a system of size $N$ is $\sim e^{-(2N)^2}$; this corresponds to a short relaxation time $\tau_{min} \sim 1/(2N)^2$.

\begin{figure*}[th!] 
\includegraphics[width=\textwidth]{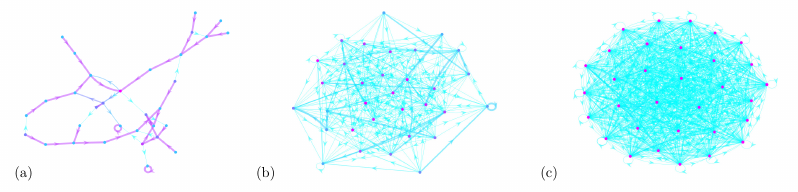}
  \caption[Figure 3]{\label{fig2} Illustrative networks at with $N=32$ at temperatures (a) $\epsilon/\epsilon_c = 10^{-2.4}$; (b) $\epsilon/\epsilon_c = 10^{-0.6}$; (c) $\epsilon/\epsilon_c = 10^{0.6}$. Edges are defined for matrix elements $M_{ab} > m_0 = 10^{-3}$ and are coloured based on their weight. Nodes are coloured based on their out-degree.
  }
\end{figure*}
\begin{figure}[tb!] 
\includegraphics[width=0.8\columnwidth]{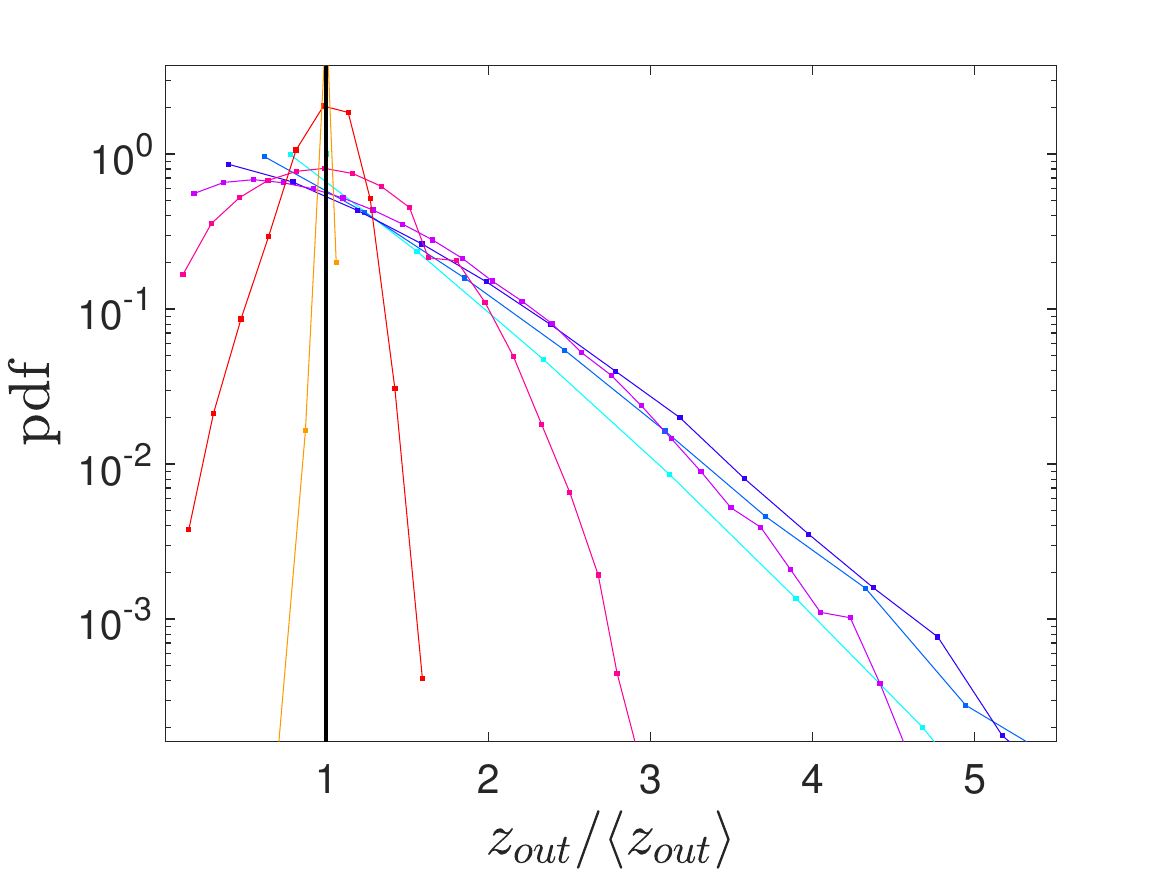}
  \caption[Figure 4]{\label{fig3} The distribution of node out-degree at $N=128$ and various $\epsilon$ from $\epsilon/\epsilon_c = 10^{-2.4}$ (cyan) to $\epsilon/\epsilon_c = 10^{1.2}$ (orange). The distribution shows a transition from being point-like for $\epsilon>\epsilon_c$ to having an exponential tail at large values at smaller $\epsilon$.
  }
\end{figure}

Hidden in Figs. 1ab is the behaviour of $\PP(|\lambda|)$ near $|\lambda|=1$, shown in Fig. 1c. Surprisingly, this also shows divergent behaviour: at the smallest $\epsilon$ we find $\PP(\al)\sim 1/[(1-\al)\log(1/(1-\al))^{5/2}]$ over ten orders of magnitude. Assuming this form, the largest magnitude eigenvalue is $|\lambda| \approx 1-e^{-(2N/3)^{2/3}}$, aside from the Perron-Frobenius eigenvalue $|\lambda_{PF}|=1$. Hence the system has a huge relaxation time $\tau_{max} \sim e^{(2N/3)^{2/3}}$, diverging exponentially fast in the system size. For intermediate $\epsilon < \epsilon_c$, our data does not resolve the details of this divergence with complete confidence, but it is clear that $\log\tau_{max} \sim N^\alpha$ for some $\alpha>0$. This generic presence of large relaxation times in our model is the main result of this work. 

{\blue We can also look at the structure of the spectra in the complex plane, shown in Fig. \ref{figspec} for 1000 matrices of size $N=64$, and several values of $\epsilon$. Aside from the disk expected from random matrix theory, and lone outliers at $\lambda=1$, we note a key feature: there is a bicycle spoke structure that develops at small $\epsilon$. The spokes are situated along roots of unity $e^{i k \pi/n}$, for $n=2,3,4,\ldots$. }


{ In the present work we will refer to $\epsilon_c$ as a phase transition. This term is used somewhat loosely, since discontinuous behavior of observables may only occur in the asymptotic large $N$ limit (taken jointly with $\epsilon \to \epsilon_c$). However, we observe smooth changes with $N$ even at the modest values of $N$ that we considered.}



Note that the specific forms of the eigenvalue tails near $\lambda=0$ and $|\lambda|=1$ are empirical fits; similar tails have been observed before in the spectra of random symmetric matrices with heavy-tailed element distributions \cite{Kuhn15,Kuhn15a}. \\

\begin{figure*}[th!] 
\includegraphics[width=\textwidth]{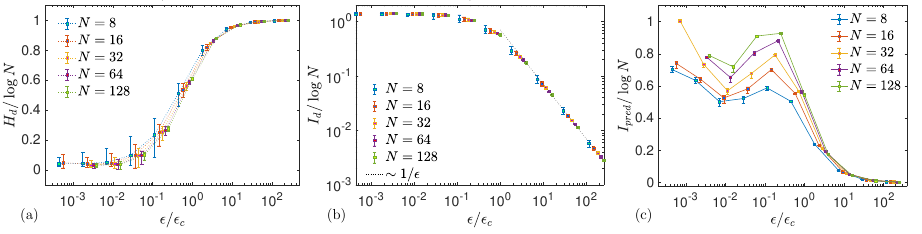}
  \caption[Figure 5]{\label{fig4} Information-theoretic quantities. (a) The Shannon entropy rate of state visitation sequences in the Markov chain at indicated $N$; (b) Information rate, showing its increase as $\epsilon$ is decreased from large values; (c) Predictive information, peaking at an intermediate $\epsilon$. Error bars in (a),(b) correspond to 20${}^{th}$ and 80${}^{th}$ percentile ranges over 1000 distinct Markov chains (samples) at each parameter value, while error bars in (c) correspond to true measurement errors from 300 distinct Markov chains (samples) at each parameter value.
  }
\end{figure*}

\subsection{Network interpretation}

To illustrate how the transition matrices change as $\epsilon$ is varied, it is useful to consider the directed network formed by considering states as nodes and transitions as directed edges. In our model, all states are connected with some probability, thus the network is always fully connected: over a long enough time-scale, transitions between any states are possible. However, the transition elements can vary over a wide range. We thus construct the networks with edges above a threshold $m_0$, which corresponds to choosing a maximum observation time-scale. {\blue Very roughly, this corresponds to the network observed on times less than $\sim 1/m_0$. } We set $m_0 = 10^{-3}$ in what follows. Typical networks for $N=32$ are illustrated in Figure \ref{fig2}, where a spring-based force algorithm was used to determine the vertex positions. Edge thicknesses and colours are determined by their weight. 

Several features are apparent. At high temperature, the network is nearly fully connected, and therefore has many loops. Statistically, vertices are equivalent. As the temperature is lowered, the graph becomes sparser. Slightly below the critical temperature, loops of various scales are present, thus allowing nontrivial temporal behaviour on a range of scales. Finally, at the lowest temperature studied, the network is very nearly a tree, which thus will lead to a trivial dynamic behaviour. 

At low temperature, the network depends on the threshold. If the threshold were to be lowered, the networks would become denser, revealing rare transitions between states. At the lowest temperatures, the very wide range of timescales implies that these networks will continue to have structure as the threshold is made ever smaller. 


The networks can be analyzed with standard measures \cite{Newman18}. For example, the out-degree of a node is the number of edges leaving it. Its distribution, shown in Fig. \ref{fig3} for $N=128$, clearly shows the transition: at temperatures above $\epsilon_c$, it is nearly a Dirac $\delta-$function peak at its average value, since the network is nearly fully connected. It broadens continuously as $\epsilon$ decreases, giving an approximately exponential tail of well-connected nodes. \\

We also investigated the node centrality (not shown). For $\epsilon>\epsilon_c$, the distribution of centrality is peaked around a typical value, while for smaller $\epsilon$ it develops a tail of large centrality nodes, and a weakly divergent tail of nodes with small centrality. \\

\section{Applications}

{\blue We now consider several applications of the random Markov ensemble. First, we consider the sequences produced by the Markov model in the light of information theory. We will show that the transition identified above has a signal in information-theoretic quantities, namely entropy and predictive information \cite{Bialek01,Bialek01a}. Then, we extend the Markov ensemble to an ensemble of Hidden Markov models, which are extensively used in sequence analysis and neuroscience. Finally, we apply the Markov ensemble to neural data, and show that several measured quantities can be quantitatively predicted from the random model. }

\subsection{Information theory \& complexity}

Since we study Markov models as complex systems, we can also look at measures of complexity using information theory. {\blue In this section, we will consider three quantities: the Shannon entropy, whose expression for stationary Markov models we review; the predictive information of \cite{Bialek01,Bialek01a}, whose construction we review; and Zipf's law. }

\subsubsection{Shannon entropy: } {\blue The Shannon entropy $H = - \langle \log \PP(X) \rangle$ of a probability distribution quantifies the amount of information gained upon measurement of a random variable }\cite{Cover99} (We define the entropy in base $e$). We can apply this to the sequences of states visited by a Markov model, for example $X = (A B A A D D E A B C \ldots)$, etc., which is an element in the space of sequences. Labelling the sites visited in a sequence as $x(1), x(2), \ldots$, we write $H(t)$ for the entropy of a sequence of length $t$. {\blue Consider a discrete and stationary Markov process. For simplicity, we assume that the initial conditions are sampled from the stationary distribution $\pi_a$, which is assumed to exist. }Then we have
\eq{
H(t) & = - \langle \log \PP(x(1),x(2),\ldots,x(t)) \rangle \notag \\
 & = - \langle \log \big[\PP(x(1)) \PP(x(2)|x(1)) \cdots \PP(x(t)|x(t-1)) \big] \rangle \notag \\
 & = - \sum_a \pi_a \log \pi_a - (t-1) \sum_{a,b} \pi_b M_{ab} \log M_{ab} \notag \\
 & = H_\pi + (t-1) H_d \label{H}
}
where we used the Markov property to expand the probability of a sequence, and we defined the entropy $H_\pi$ of the stationary distribution $\pi_a$. At large times $H(t)$ is dominated by the {\it entropy rate}
\eq{
H_d = \lim_{t \to \infty} \ffrac{1}{t} H(t)
}
 {\blue If the Markov chain is irreducible but the initial conditions are not sampled from the stationary distribution, then there will be transient corrections to \eqref{H}. }

$H_d$ is plotted in Fig. \ref{fig4}a over the full range of temperature and system size $N$ \footnote{For simplicity we consider only ergodic samples. Ergodicity can break down at small $\epsilon$ due to floating-point issues.}. When $H_d$ is normalized by its maximal value $\log N$, and temperatures are normalized by $\epsilon_c(N)$, then the curves collapse at large $\epsilon$ and also at small $\epsilon$, up to small corrections logarithmic in $N$ (no power of $\log N$ produces a collapse at all $\epsilon$, indicating a crossover between two functional forms). The normalized entropy rate tends to unity at large temperature, indicating that transition sequences are indistinguishable from random noise, while at the lowest temperatures it is very nearly 0, indicating that sequences are becoming nearly deterministic. 

Equivalently, we can define the information rate $I_d = (\log N - H_d)/\log 2$ such that $I_d$ is the typical amount of information contained in the probability distribution per entry of the sequence, in bits \cite{Parisi88}. {\blue Again, this holds for stationary conditions. }This is shown in Fig. \ref{fig4}b on logarithmic axes, highlighting the increase in information rate as $\epsilon$ is decreased. The dotted line shows the function $I_d/\log N \propto \epsilon_c/\epsilon$ which tracks this increase for $\epsilon \gtrsim \epsilon_c$. 

The error bars shown in Figs. \ref{fig4}ab correspond to 20${}^{th}$ and 80${}^{th}$ percentile ranges over 1000 distinct Markov chains (samples) at each parameter value \footnote{The true error bars in the measurements are then smaller by a factor of $\sim 1/\sqrt{1000} \sim 1/32$.}. We see that these bars decrease as $N$ increases, so that the normalized entropy rate $H_d/\log N$ is a {\it self-averaging} quantity: it doesn't fluctuate from sample-to-sample, in the large size limit. This is a partial {\it a posteriori} justification for the lognormal distribution of matrix elements, since it implies that $\epsilon$ controls the entropy, at least for large systems. Thus, for example, if we measure the normalized entropy rate in some system under some dynamic process, then we can infer how $\epsilon$ varies. 

\subsubsection{Complexity: } Entropy distinguishes order from randomness, and thus captures the emergence of structure in Markov models, but it does not necessarily measure complexity. Intuitively, from Fig. \ref{fig2}, one expects that models at very low temperature are not complex because they are too deterministic, while models at high temperature are not complex because they are completely random. Can this intuition be quantified? 

Grassberger \cite{Grassberger86} suggested that the complexity of a sequence can be measured by how quickly the entropy rate $H[x(1),x(2),\ldots,x(t)]/t$ attains its asymptotic value. This notion was formalized by Bialek, Nemenman, and Tishby in \cite{Bialek01,Bialek01a} with the notion of predictive information. For a stationary process one asks how well the `past' $(x(-t),x(-t+1),\ldots,x(-1))$ predicts the `future' $(x(0),x(1),\ldots,x(t-1))$. This is quantified by the mutual information \cite{Cover99} between past and future, $I(t,t') = H(t) + H(t') - H(t+t')$ where we are using the fact that the process is stationary, so that this depends only the length $t$. Bialek et al then define the predictive information as 
\newcommand{\pred}{\text{pred}}
\eq{
I_{\pred}(t) = \lim_{t' \to \infty} I(t,t')/\log 2,
}
which thus measures how well the past $t$ states predict the entire future trajectory of the system. 

Since $H_d$ is the asymptotic extensive part of $H(t)$, the latter can be written $H(t)= t H_d + H_1(t)$ where $H_1(t)$ is the nonextensive part. Clearly dependence on $H_d$ drops out of $I_{\pred}$, showing the insufficiency of $H_d$ to measure predictive power, and thus complexity, in this theory (This statement holds for general stationary processes, not necessarily Markov ones that follow  \eqref{H}). Consistent with the arguments of Grassberger, complexity is measured not by the asymptotic state but by the approach to it.

For a discrete and stationary Markov process, {\blue under the same assumptions as \eqref{H},} it follows from \eqref{H} that
\eq{
I(t,t') = H_\pi - H_d \label{I}
}
which is independent of both $t$ and $t'$, a non-generic property \cite{Bialek01,Bialek01a}. The predictive information is plotted in Fig. \ref{fig4}c. We confirm that it has a peak at intermediate $\epsilon \sim 0.1 \epsilon_c$, although this precise location is not resolved in our data. This furthermore supports the notion that the high and low-temperature phases are separated by a phase transition, rather than just a cross-over.

Curiously, we find that $I_\pred$ also increases as $\epsilon$ becomes very small. The interpretation of this result we leave for future work. 

\subsubsection{Zipf's law: } 
\begin{figure}[tb!] 
\includegraphics[width=0.8\columnwidth]{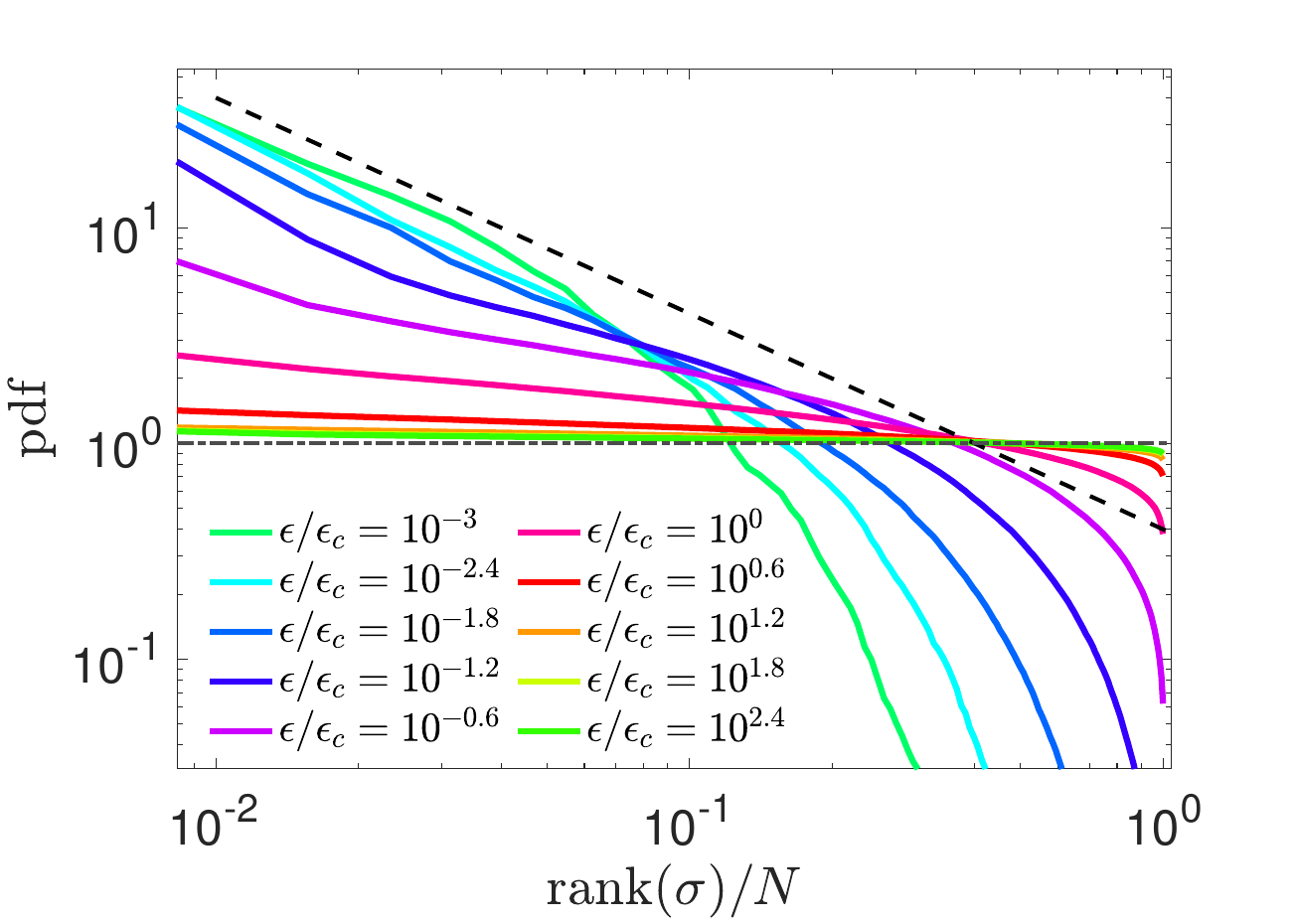}
  \caption[Figure 6]{\label{fig5} Probability distribution of the rank of states (denoted by $A$ in the axis label) for $N=128$. Zipf's law is a power-law relation $\PP(r) \sim 1/r$, as indicated by the dashed line. 
  }
\end{figure}

Finally, often discussed in the context of sequence complexity is Zipf's law \cite{Cancho03,Corominas-Murtra10,Corral15,Schwab14,Zipf13}. {\blue Consider the sequence of states visited by the Markov model. Zipf's law }states that when states $A$ are ranked in terms of decreasing frequency of appearance, their probability is inversely proportional to their rank: $\PP(r(A)) \sim 1/r(A)$, where $r(A)$ is the rank of the state $A$. Zipf's law is found to hold to a good approximation in human language texts (when `states' are words), as explained by various theories \cite{Cancho03,Corominas-Murtra10,Corral15,Schwab14,Zipf13}. This quantity is shown for our $N=128$ data in Fig \ref{fig5}, on logarithmic axes. For $\epsilon>\epsilon_c$, the curve is nearly flat: all sites are visited with the same frequency. Below $\epsilon_c$, the curve develops a nontrivial shape, including approximate power-law regions, and for very small $\epsilon$ one finds Zipf's law, indicated by the dashed line. \\

%
%
%

\subsection{Hidden Markov models}

In most dynamical systems, we have only incomplete knowledge of the detailed configuration of the system components. Often we can consider that the system undergoes a Markovian evolution, as above, but we only have access to a sequence of observables, which are outputted as the internal system changes state. This leads to the definition of a Hidden Markov model \cite{Rabiner86,Eddy04}: we have (i) a set of internal `hidden' states, $\{1,2,\ldots,N\}$ with an internal transition matrix $M_{ab}$, as above; and (ii) in addition, a set of observable states $\{1,2,\ldots,T\}$ (distinct from the hidden states), and an emission matrix $O_{aB}$ giving the probability that when the system is in hidden state $a$ it outputs observable $B$. It is supposed that one observable symbol is outputted between each transition of hidden states.

For example, while an infant may only be able to produce a small number of distinguishable noises, its hidden mental state may be significantly richer. Hidden Markov models (HMMs) have found application in various domains, particularly in bioinformatics \cite{Eddy04,Durbin98}, natural language processing \cite{Juang91}, and, most recently, in neuroscience \cite{Ou15,Vidaurre17,Vidaurre18,Stevner19,Goucher-Lambert19}. HMMs are also the simplest level of the Chomsky hierarchy when applied to probabilistic grammars, and thus serve as toy models for syntactic structure \cite{Chomsky02,Carnie13,DeGiuli19}. 


All of our earlier considerations apply to the sequence of hidden states visited by a HMM. When we add an emission matrix, then we can also investigate quantities that depend only on the observed states. {\blue The same arguments that led us to consider a lognormal distribution for the elements of the Markov matrix also apply to the emission matrix. We introduce a primitive matrix $P$ and write $O_{aB} = P_{aB}/\sum_C P_{aC}$. We consider the matrix $P$ to be drawn from a lognormal distribution}, now with a `surface' temperature $\epsilon_s$, which we fix to $\epsilon_s = 0.01$. We also fix $T=12$. 

{\blue We will investigate the sequences produced by the HMM both in terms of their entropy, and in terms of their complexity. }

\subsubsection{Entropy: } We consider first the Shannon entropy rate of observed sequences, $H_s$. There is no simple expression for the entropy rate in terms of $M$ and $O$, although rigorous algorithms exist for its estimation \cite{Jurgens20}. Here we estimate $H_s$ by sampling, using the method of Grassberger to eliminate sampling biases \cite{Grassberger03}. First we define the $k-$gram surface entropy rate $H_s{(k)} = \ffrac{1}{k} H(o(1),o(2),\ldots,o(k))$ where the $o(j)$ are the observable symbols. The asymptotic entropy rate can be estimated directly by the limit $H_s{(k)}$ as $k \to \infty$, but it is also estimated by the differential entropy rate 
\eq{
\delta H_s{(k)} = (k+1) H_s{(k+1)} - k H_s{(k)},
}
and the latter converges faster \cite{Schurmann96}. We estimate these entropy rates by constructing 300 different Markov models at each value of $N$ and $\epsilon$, and sampling a sequence of total length $\sim 10^5$ for each sample, at each parameter value. This allows us to estimate $H_s{(k)}$ for $k \leq 4$ \footnote{As a crude estimate, since the $k^{th}$ entropy rate has a phase space of size $T^k$, we are limited to $T^k \ll 10^5$, or $k < 4.6$.}. 

\begin{figure}[tb!] 
\includegraphics[width=0.8\columnwidth]{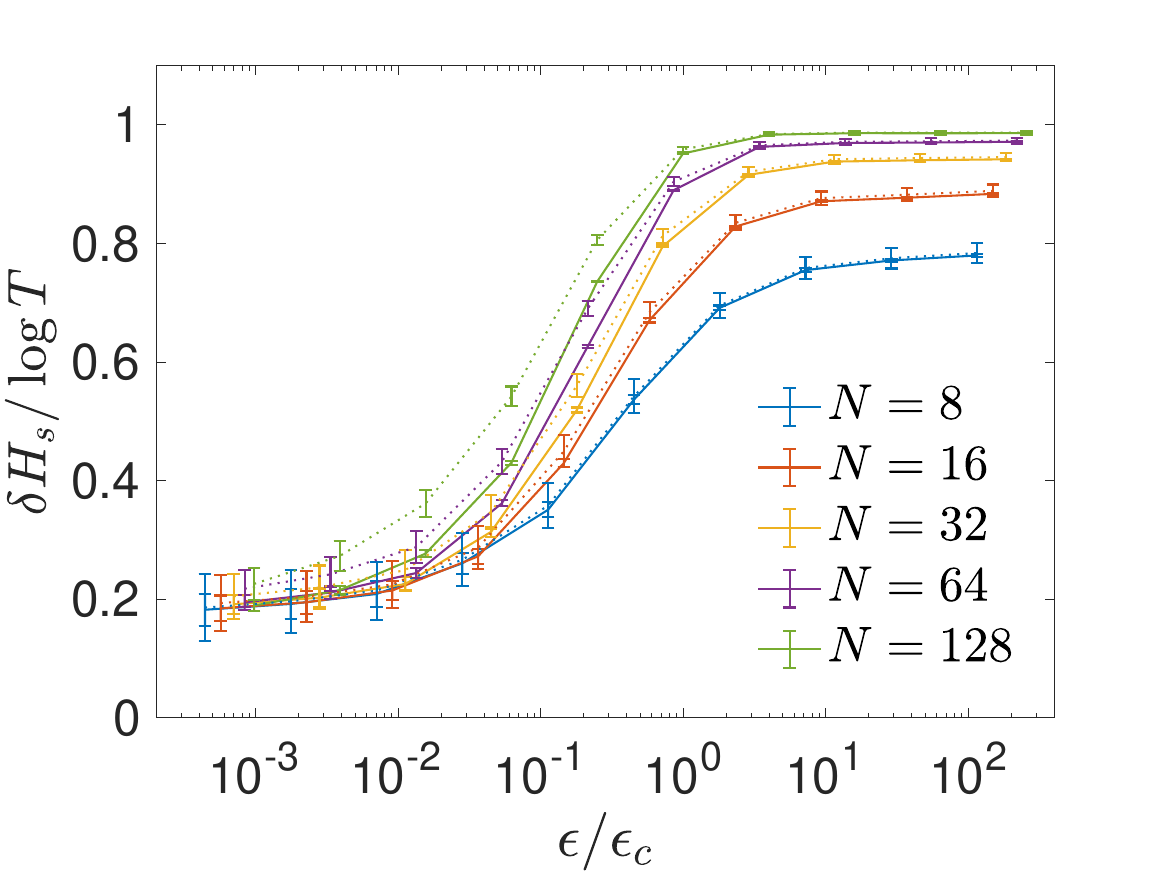}
  \caption[Figure 7]{\label{fig6} Differential surface Shannon entropy rates $\delta H_s(k)$ for Hidden Markov model with  $k=3$ (solid) and $k=2$ (dotted), at indicated $N$. Error bars correspond to 20${}^{th}$ and 80${}^{th}$ percentile ranges over 300 distinct Markov chains (samples) at each parameter value. Here $T=12$ and $\epsilon_s = 0.01$. 
  }
\end{figure}

The resulting $\delta H_s{(3)}$ and $\delta H_s{(2)}$ are plotted in Fig.~\ref{fig6} as solid and dotted lines, respectively. First, we notice that for $N \leq 16$ the two differential entropy rates are very nearly equal, and thus close to their asymptotic value everywhere. For larger $N$, the rates are nearly converged for $\epsilon > \epsilon_c$ but display some finite-$k$ effects at smaller $\epsilon$. These effects in fact indicate correlations in the sequences \cite{Bialek01}, studied below. Here we focus on the systematic trends with $\epsilon$ and $N$. 

Consider a sequence outputted by any given HMM. It has two sources of randomness: that in $M$, and that in $O$. In our case, since $\epsilon_s = 0.01$ is relatively small, the $O$ matrix is very sparse: given the hidden states, the observed sequence is nearly deterministic. Variation in  $\delta H_s$ is thus due to changes in the physics of the hidden states. 

Consider first $\epsilon > \epsilon_c$. If the outputted sequence is completely random noise, then we will have $\delta H_s \approx \log T$, as indeed we find for large enough $N$. As $N$ is lowered, $\delta H_s$ also decreases. The interpretation of this regime is as follows: even if the hidden state physics is completely random, outputted sequences can have structure simply because $O$ need not output each observed symbol with the same frequency. For example, if the symbol $A$ is over-represented in the output of $O$, then this will impact the observed entropy rate, even for uniformly random hidden states. {\blue This is why $\delta H_s$ can be small even when $\epsilon > \epsilon_c$. However, there is a competing averaging effect: when the hidden states are sampled uniformly, each $O$ element only contributes in the form $\frac{1}{N} \sum_{a} O_{aB}$, which is averaged over the hidden states. When $N$ becomes large, this is subject to the law of large numbers and this becomes equal to $\langle O_{aB} \rangle$, which is then independent of $B$ (in our model). Thus, finally, for large $N$ and $\epsilon > \epsilon_c$, the entropy becomes again that of a uniform random variable, and tends to $\log T$ as $N \to \infty$, as observed. }


\begin{figure}[tb!] 
\includegraphics[width=0.8\columnwidth]{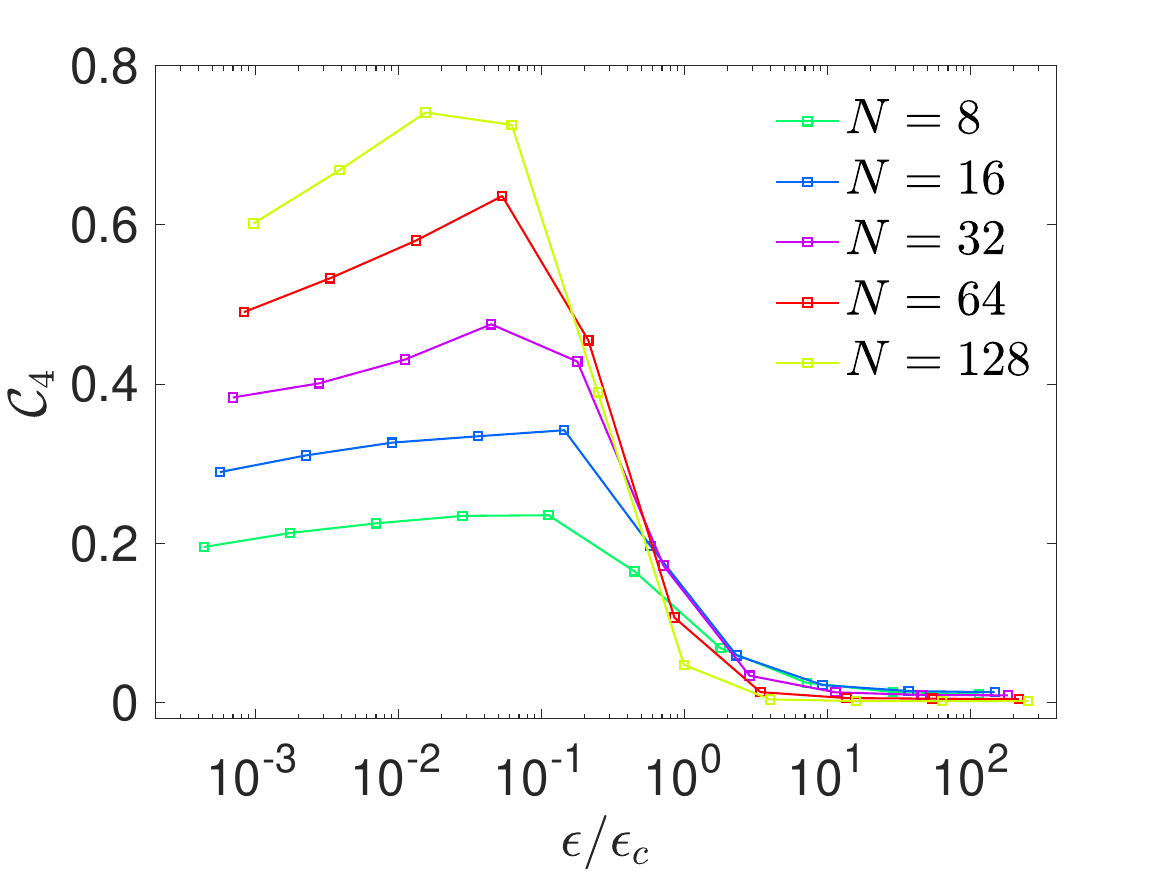}
  \caption[Figure 8]{\label{fig7} Complexity measure for observable sequences in Hidden Markov model,  with $k=3$. Consistent with the predictive information measure for the hidden states, this peaks at an intermediate temperature below $\epsilon_c$. 
  }
\end{figure}

As we decrease $\epsilon$, the observed entropy rate always drops substantially from its value at $\epsilon_c$: the physics of hidden states always plays an important role. We notice that at sufficiently small $\epsilon$, the curves for different $N$ collapse on top of one another. The asymptotic small $\epsilon$ entropy rate depends only on $T$ and $\epsilon_s$. In this regime, the hidden state evolution is effectively bottlenecked by $O$: even if it is completely deterministic, the observed sequences have randomness simply due to that in $O$. 

\subsubsection{Complexity: } We can also investigate the complexity of observable sequences. The full $I_\pred$ is numerically challenging to obtain, but following the above discussion we can define an observable that measures the approach to the asymptotic entropy rate. As a simple measure, we define a complexity $\mathcal{C}_k = -H_s(k) +\ffrac{k}{k-1}H_s(k-1)$, plotted in Figure \ref{fig7}. This combination of entropies eliminates the contribution to $H_s(k)$ from the asymptotic entropy rate, and thus depends only on the rate at which this asymptotic rate is approached. Consistent with $I_\pred$ defined above, we find that $\mathcal{C}_k$ is non-monotonic, with a peak approximately at $\epsilon \sim 0.1 \epsilon_c$. 

A consequence of these analyses is that for a HMM, many properties of the observable sequences mirror those of the underlying hidden sequences. This is useful, because the hidden sequence is Markovian, and therefore easier to analyze {\it a priori}. In the usual scenario where only observable sequences are known, this is tempered by the fact that the hidden sequence must be reconstructed from the observations. Fortunately, efficient algorithms and toolboxes exist for this process \cite{Rabiner86,Vidaurre18}. \\

\begin{figure}[th!] 
\includegraphics[width=0.8\columnwidth,viewport=70 70 520 400,clip]{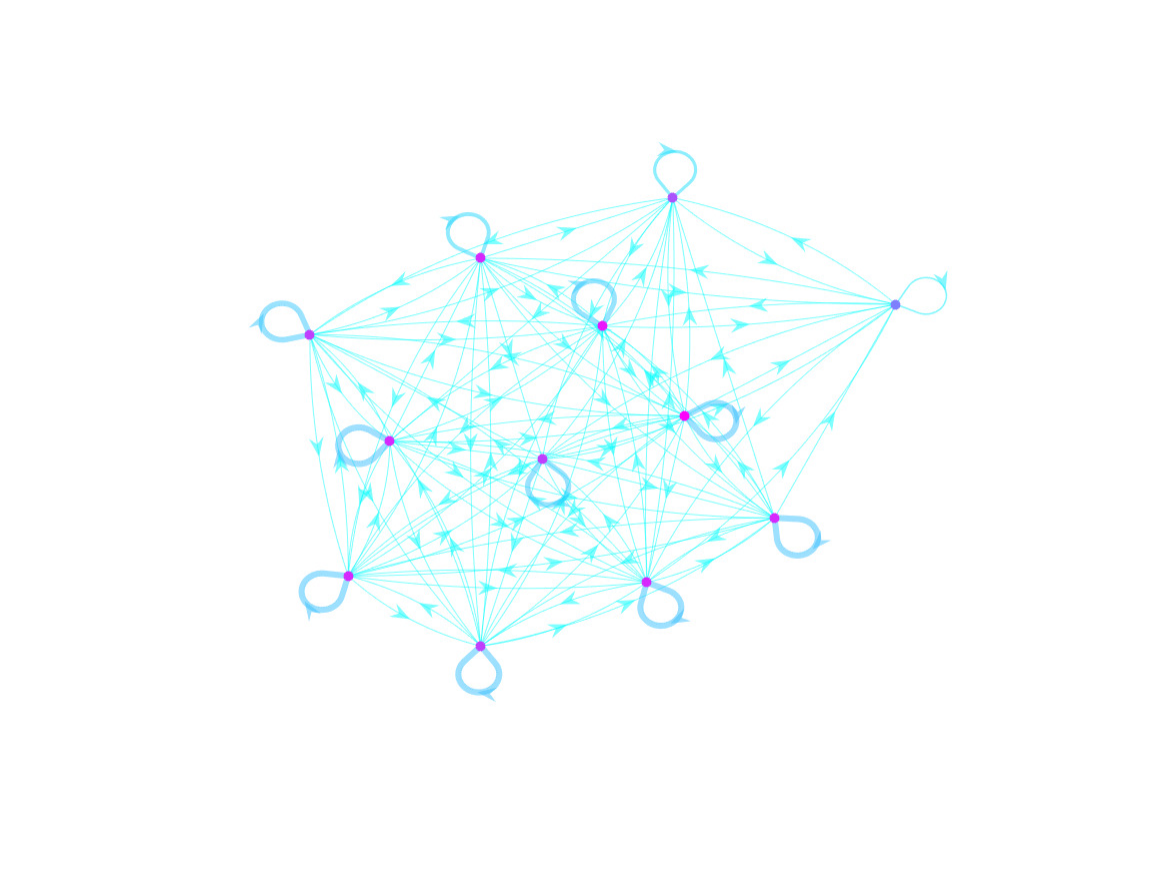}
  \caption[Figure 9]{\label{fignetwork} Representative network for the Hidden Markov model applied to neural data.
  }
\end{figure}

\begin{figure*}[th!] 
\includegraphics[width=\textwidth]{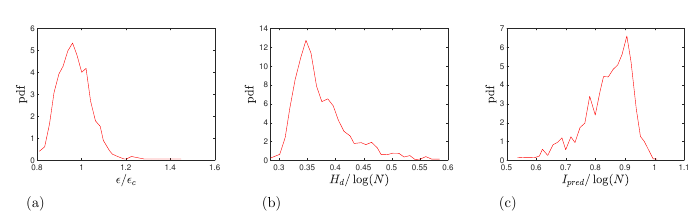}
  \caption[Figure 10]{\label{fig8} Hidden Markov model applied to neural data. Probability distributions of (a) temperature; (b) Shannon entropy rate of hidden sequences; (c) predictive information of hidden sequences.
  }
\end{figure*}
\begin{figure*}[th!] 
\includegraphics[width=\textwidth]{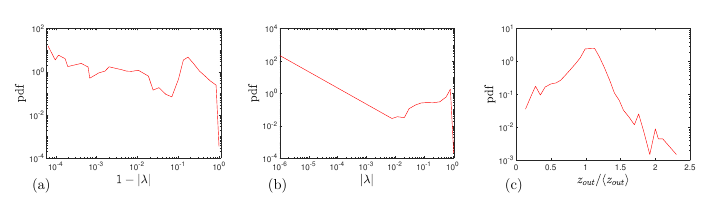}
  \caption[Figure 11]{\label{fig9} Hidden Markov model applied to neural data. (a,b) Eigenvalue spectra; (c) out-degree distribution.
  }
\end{figure*}

\subsection{Hidden Markov model applied to neural data}

The explosion in availability of neural fMRI data requires tools that can reduce the dimensionality of the data, and connect the data to functionally significant `modes' of neural activity \cite{Lurie20}. Hidden Markov models are well-suited to this task, and have been extensively applied in neuroscience \cite{Ou15,Vidaurre17,Vidaurre18,Stevner19,Goucher-Lambert19,Vidaurre21}.

To test the extent to which the random HMM discussed above applies to these experimental data, we obtained the HMM transition matrix $M$ and numerous hidden-state sequences from \cite{Vidaurre17}, which itself uses neural data from the Human Connectome project \cite{Smith13}. The whole-brain fMRI data is taken from the resting-state activity of 820 human subjects. $N=12$ states were inferred at the group-level, using multivariate Gaussian distributions applied to the raw fMRI data. After determining the states, the transition matrices $M$ were inferred at the subject level. {\blue Pseudocounts were added to the state transition counts, following Laplace's rule of succession \cite{Barton14}.} Approximately 80\% of subjects used all 12 states, while 20\% used 11 states, and a handful used 8-10 states. In our subsequent analysis, we defined and used an $N$ value for each subject.

A typical state-by-state network from this data is shown in Fig.\ref{fignetwork} (no threshold was used). Comparing with the networks in Fig.\ref{fig2}, a conspicuous feature here is that the self-transitions $A \to A$ are much more probable than transitions between states. This possibility was not taken into account in the random model, but its consequences will be deduced below.

{\blue Note that although this data was fitted to a Hidden Markov model, we will analyze only properties of the hidden states, so that we work in the context of the original Markov ensemble. }

First, we naively analyze the HMMs in the context of the random model. For each subject, we measure the heterogeneity $h$ of its $M$ matrix and infer a temperature $\epsilon \equiv 1/(4[-1+\sqrt{1+h}])$, the entropy rate $H_d$, using \eqref{H}, and $I_\pred$, also derived from \eqref{H}. The probability distributions of these quantities, rescaled as in earlier sections, are shown in Fig.\ref{fig8}. All of these quantities are sharply peaked around typical values, compared to the vast region explored, for example, in Fig. \ref{fig4}, which extends over 5 orders of magnitude in $\epsilon/\epsilon_c$. In particular, the most likely temperature is very near the critical one, a striking result given its status as a phase transition, and the peak in complexity near $\epsilon_c$ (Fig.\ref{fig4}c). The mean values are $\epsilon/\epsilon_c \approx 0.97$, $H_d/\log N \approx 0.375$, and $I_\pred/\log N \approx 0.84$. However, if we naively try to match these values with the random model, the comparison clearly fails: for example $\epsilon \approx \epsilon_c$ would imply that $H_d/\log N \approx 0.6$, significantly higher than the observed value. How can we interpret this?

The key is the feature already noticed in Fig.\ref{fignetwork}, namely that self-transitions are much more probabe than transitions between states. Quantitatively, we find that the probability of a self-transition is $p \approx 0.75$, with fluctuations $\sim \pm 0.06$. For simplicity consider that all self-transitions have this same probability $p$. Then we can decompose the entropy rate as follows:
\eq{
H_d & = - \sum_{a,b} \pi_b M_{ab} \log M_{ab} \notag \\
& = - \sum_{b} \pi_b p \log p - \sum_{b} \sum_{a \neq b} \pi_b M_{ab} \log M_{ab} \notag \\
& = - p \log p - \sum_{b} \sum_{a \neq b} \pi_b (1-p) \tilde M_{ab} \log [(1-p) \tilde M_{ab}] \notag \\
& = [- p \log p - (1-p) \log (1-p)] \notag \\
& \qquad - (1-p) \sum_{b} \sum_{a \neq b} \pi_b \tilde M_{ab} \log \tilde M_{ab} \notag \\
& = H_{\text{Ber}}(p) + (1-p) \tilde H_d \label{Hdcond}
}
where $H_{\text{Ber}}(p)$ is the entropy of a Bernoulli process with probability $p$ and $\tilde M_{ab} = M_{ab}/(1-p)$ is the conditional probability of the transition $b \to a$, given that this is not a self-transition. Thus the entropy rate is decomposed into the entropy for the decision whether or not to transition between states, and the remaining entropy of the conditional distribution described by $\tilde M_{ab}$. The crucial point is that the latter contribution is weighted by the small factor $1-p \approx 0.25$. Thus if the conditional process is considered a random process with $\tilde H_d/\log N \approx 0.6$, we expect $H_d/\log N \approx (H_{\text{Ber}}(0.75)/\log 12) + 0.25 \times 0.6 = 0.375$, which is exactly what we find.

We can apply a similar analysis to $I_\pred$. If $\epsilon=\epsilon_c$, then from Fig. \ref{fig4}c we estimate $\tilde I_\pred/\log N \approx 0.4$ for the random model. Using \eqref{I} this implies $H_\pi/\log N = \log(2) \times 0.4 + 0.6 = 0.88$, which, again using \eqref{I}, then implies $I_\pred/\log N = (0.88 - 0.26)/\log(2) = 0.89$. This is remarkably close to the measured mean value $0.84$. 

Altogether, if we know the value of $p$, and assume $\epsilon=\epsilon_c$, then we can quantitatively predict the values of the entropy rate and the predictive information. This was unexpected, because our random model contains only two parameters, namely the number of hidden states $N$ and the temperature $\epsilon$. Naively, one might imagine that many parameters would be needed to capture the properties of the HMM. This is especially so in the present context since it was shown in \cite{Vidaurre17} that the HMM contains information on functional modes of neural activity (such as motor network {\it vs} language network {\it vs} visual network), which are known to exhibit considerable spatial and temporal complexity. 

To confirm that these results are not a coincidence, we can check that other properties of the random model hold for the neural data. To this end we measure the spectra of the transition matrices and the degree distribution of the networks. As earlier, we plot the eigenvalue density v.s $1-|\lambda|$ and v.s $|\lambda|$, and the distribution of $z_{out}/\langle z_{out} \rangle$, in Fig \ref{fig9}abc, respectively. Although noisy, these distributions have key features of Fig 1c, Fig. 1b, and Fig. \ref{fig3}, respectively. In particular: $\PP(|\lambda|)$ has a tail of small values, corresponding to short relaxation times, as well as a tail of values close to 1, corresponding to long relaxation times, and $\PP(z_{out})$ is peaked around unity with the beginnings of an exponential tail at large values. 

Let us note that the parameter $p$ is related to the sampling rate of the data: if the decision whether or not to change state is a Bernoulli process, then the number of time steps $\Delta n$ before a true transition is a geometric random variable, with mean $1/(1-p)$. These parameters are thus equivalent. If the sampling rate is $f$ (in Hz) then the typical time (in s) between transitions is $\Delta t = \langle \Delta n \rangle/f = 1/(f(1-p))$. 

Finally, let us comment on the surprising finding $\epsilon \approx \epsilon_c$, that is, subjects lie at the phase transition at which the large$-N$ random matrix theory result breaks down. Since long relaxation times exist in the entire low-temperature phase, there is no simple {\it a priori} reason to expect neural activity to be right at this transition. However, since complexity peaks near the transition (Fig. \ref{fig4}c), this could be explained if brains are driven towards a state of maximal complexity.


These results lend support, at the whole-brain level, to the hypothesis that the brain operates in a critical regime, adding mathematical evidence to previous studies on small populations of individual neurons, and EEG \cite{Arcangelis06,Beggs08,Deco12,Hesse14,Tkacik15,Fontenele19,Fosque21}. \\


\section{Conclusion}

 The analysis of complex systems is often limited by the lack of a nontrivial null model to which one can compare data. One approach to this problem is to consider the ensemble of all models of a given class, for example discrete Markov models of a given size. The ensemble is generally vast, including all possible models to which the data could be fitted, and cannot be completely characterized if the system under consideration has a high complexity. However, if one can identify a few key parameters that control variation of interesting observables in the ensemble, then one can consider this {\it random model} as a null model. One can then look for phase transitions in the ensemble, with the aim of mapping out the general behaviors available within the class of model. This programme has been advocated as a general strategy for the study of complex systems \cite{Parisi99}, and has been attempted for constraint satisfaction problems \cite{Mezard02}, for language syntax \cite{DeGiuli19,DeGiuli19a}, and for ecosystems \cite{Biroli18}, to give a few examples. 

In this work we have proposed a simple ensemble of discrete Markov models, and a straightforward extension to Hidden Markov models. As a control parameter we introduced $\epsilon$, which we call a temperature. The physical interpretation of $\epsilon$ is that it controls the dynamic range of transition matrix elements: large $\epsilon$ corresponds to a small dynamic range. In other words, when $\epsilon$ is large the various transitions have similar probabilities, while when $\epsilon$ is small, these probabilities can be vastly different. We have shown that the $\epsilon$ plays a key role in the behavior of the Markov models: there is a critical value $\epsilon_c$ such that large-$N$ random matrix theory results hold for $\epsilon>\epsilon_c$, but not otherwise. In the low-temperature phase, the Markov models generically have a broad spectrum of relaxation times, without any fine-tuning. The critical point $\epsilon_c$ thus acts as a phase transition between a simple high-temperature phase, and a more ordered low-temperature phase. In our ensemble the value $\epsilon_c = 1/(2 \log (N+1))$ depends only on the number of states in the Markov model. We have furthermore shown that $\epsilon$ controls the entropy of state-transition sequences. 

The utility of this ensemble was explicitly shown using human fMRI data, where we could quantitatively characterize the values of entropy and predictive information found in a HMM fit to fMRI data. { The latter measures how informative observations of the `past' are with respect to predicting the future, and is an essentially unique measure of complexity \cite{Bialek01a}. }


Owing to the generality of Markov models, this work fits into a number of larger problems. First, Hidden Markov models are the simplest class of probabilistic grammars, which are organized by the Chomsky hierarchy \cite{Chomsky02,Carnie13,DeGiuli19}. The next level of complexity in grammars are the context-free grammars, considered as an ensemble in \cite{DeGiuli19, DeGiuli19a,Nakaishi21}. These are related to syntactic structure in both natural and computer languages. A similar phase transition to that discussed here was found in \cite{DeGiuli19, DeGiuli19a}, but without any interpretation in terms of random matrix theory. As discussed therein, the phase transition can be interpreted in terms of breaking of permutation symmetry, an interpretation that also holds in the present case. {\blue Briefly, at high temperature, all the symbols are statistically equivalent: the permutation symmetry among them is preserved. As temperature is lowered through $\epsilon_c$, the transition rates span a broad enough dynamic range that the symbols can then be distinguished. This is precisely what is measured by the Zipf plot, Fig. \ref{fig5}.} This connects to the standard Landau paradigm of phase transitions in terms of symmetry breaking. 

{\blue Second, the low-temperature regime of the Markov ensemble has matrices that are effectively sparse, thus the phase transition discussed here can be interpreted as a full-to-sparse transition of random matrices. } The behaviour of sparse random matrices, while important, is understood much more poorly than that of their dense cousins. Recent work has built theoretical tools for such study \cite{Kuhn15,Kuhn15a,Susca21}, and could be useful to theoretically derive the behaviour of the eigenvalue tails we found near $|\lambda|=0$ and $|\lambda|=1$. {\blue Eigenvalue tails have previously been predicted in models of complex networks \cite{Dorogovtsev03}, but in the symmetric case. We found that the eigenvalue spectra in the low-temperature regime show a bicycle wheel structure in the complex plane, with spokes at the roots of unity. We are not aware of any previous mention of this phenomenon. }

{\blue Third, we have considered Markov models in discrete time, but a natural generalization would be to continuous time. This would also allow a connection to stochastic thermodynamics \cite{Van-den-Broeck13}. } 

Finally, the presence of a phase transition allows one to place stochastic matrices in the phase diagram, and evaluate how near or far they are from the transition. Applied to fMRI data, we found that the data are very near the transition, supporting the brain criticality hypothesis. A natural question is whether stochastic matrices from other complex systems are also near this transition. Since the ensemble includes all discrete Markov models, it can unify the study of disparate systems, in the goals of seeking universal patterns. This may shed light on the origin and possible universality of criticality in biological systems.

{\bf Acknowledgments: } We are grateful to Anna Stavyska for her preliminary work on this project, and to a referee for a useful suggestion. EDG acknowledges support of the Natural Sciences and Engineering Research Council of Canada (NSERC), Discovery Grant RGPIN-2020-04762. DV is supported by a Novo Nordisk Emerging Investigator Award (NNF19OC-0054895) and by the European Research Council (ERC-StG-2019-850404).
\vfill 

\vfill

\bibliography{../Glasses,../language,../Biology}

\vfill 

\begin{widetext}
{\bf Appendix: From continuous-time dynamics to the master equation. }\\

Here we discuss how the transition matrix $M$ is related to the interaction matrix $\tilde M$. It is simplest to discretize time, $x$, and $\xi$. A continuum limit can be taken later, if necessary.

For example, if we discretize time, $x$, and $\xi$, then \eqref{x1} becomes 
\eq{
x_i(t+dt) = x_i(t)(1-dt) + dt\tilde M_{ij} x_j(t) + d\xi_i(t).
}
This induces a motion in phase space with a transition matrix between states $x=(x_1(t),x_2(t),\ldots,x_n(t))$ and $y=(x_1(t+1),x_2(t+1),\ldots,x_n(t+1))$ as
\eq{
M_{yx} = \sum_{\{ d\xi \}} \PP(d\xi) \prod_i \delta_{y_i - x_i (1-dt) - dt \tilde M_{ij} x_j - d\xi_i},
}
where $\PP(d\xi)$ is the probability of noise increments $d\xi=(d\xi_1,\ldots,d\xi_n)$, and the Kronecker $\delta$'s impose the dynamics, which trivially generalizes to any nonlinear extension of \eqref{x1}. Notice that $\sum_{y} M_{yx} = 1$ and each $M_{yx} \geq 0$. The phase space dynamics is given by the master equation
\eq{ \label{markov1app}
\rho(y,t+dt) = \sum_x M_{yx} \rho(x,t), 
}
where $\rho(x,t)$ is the probability of state $x$ at time $t$. If each $x_i$ has $n_x$ different states, then $M$ is an $N\times N$ matrix, with $N=(n_x)^n$.

Let us now generalize to the continuum limit in $x$ and $\xi$. $M$ becomes a function of two arguments
\eq{
M(y,x) = \int dP(\xi) \prod_i \delta(y_i - x_i (1-dt) - dt \tilde M_{ij} x_j - d\xi_i).
}
For simplicity we assume $\xi$ is Gaussian with $\langle d\xi_i\rangle=0$ and correlation matrix $C_{ij} dt = \langle d\xi_i d\xi_j \rangle$. Then
\eqs{ 
\rho(y,t+dt) & = \int dx M_{yx} \rho(x,t) \\
& =  \int dx \rho(x,t) \int dP(\xi) \prod_i \delta(y_i - x_i (1-dt) - dt \tilde M_{ij} x_j - d\xi_i) \\
& =  |1-dt + dt \tilde M|^{-1}  \int dP(\xi) \rho([1-dt + dt \tilde M]^{-1} (y-d\xi),t) \\
& =  e^{-dt (-n +\tr \tilde M) + \OO(dt^2)} \int dP(\xi) \rho(y - d\xi + dt y - dt \tilde M y + \OO(dt^2),t) \\
& =  [ 1 + dt (n - \tr \tilde M) + \OO(dt^2)] \int dP(\xi) \left[ \rho(y,t) + (dt y - dt \tilde M y - d\xi) \nabla \rho(y,t) + \half d\xi d\xi : \nabla \nabla \rho(y,t) + \OO(dt^2) \right] \\
& =  \rho(y,t) + dt (n - \tr \tilde M) \rho(y,t) + dt (y - \tilde M y) \nabla \rho(y,t) + \half dt C : \nabla \nabla \rho(y,t) + \OO(dt^2) \\
& =  \rho(y,t) + dt \nabla \cdot \big( (y - \tilde M \cdot y) \rho(y,t) \big) + \half dt \nabla \nabla : (C\rho(y,t)) + \OO(dt^2) \\
}
or
\eq{
\p_t \rho(y,t) & =  \nabla \cdot \big( (y - \tilde M \cdot y) \rho(y,t) \big) + \half \nabla \nabla : (C\rho(y,t)) 
}
which is the Fokker-Planck equation, as expected.
\end{widetext}

\end{document}